\documentclass[]{aipproc}

\layoutstyle{6x9}

\SetInternalRegister\hbadness{8000} %

\newcommand\doingARLO[2][]{%
  \ifx\mmref\undefined #1\else #2\fi
}

\def\be{\begin{equation}}
\def\ee{\end{equation}}
\def\bea{\begin{eqnarray}}
\def\eea{\end{eqnarray}}

\begin{document}

\title
{Status and prospects of di-jet production in high-energy polarized proton-proton collisions at RHIC at $\sqrt{s}=200\,$GeV} 

\author{Bernd Surrow (For the STAR Collaboration)}{
  address={Massachusetts Institute of Technology \\ 77 Massachusetts Avenue, Cambridge, MA 02139, USA},
  email={surrow@mit.edu}
  thanks={}
}

\iftrue

\fi

\copyrightyear  {2001}

\begin{abstract}
The STAR experiment at the Relativistic Heavy Ion Collider (RHIC) at
Brookhaven National Laboratory (BNL) is carrying out a spin physics
program colliding transversely or longitudinally polarized proton
beams at $\sqrt{s}=200-500\,$ GeV to gain deeper insight into the spin
structure and dynamics of the proton. These studies provide
fundamental insight into Quantum Chromodynamics (QCD).  One of the main
objectives is the determination of
the polarized gluon distribution function, $\Delta g$, through the
measurement of the longitudinal double-spin asymmetry, $A_{LL}$, for
various processes.  Inclusive hadron and jet production from polarized
pp collision data collected so far at $\sqrt{s}=200\,$GeV using the
STAR detector at RHIC have placed important constraints on $\Delta
g$. Di-jet production provides direct access to the initial parton
kinematics at leading order (LO) QCD and thus provides sensitivity to
the Bjorken-$x$ dependence of $\Delta g$. The status of the
mid-rapidity di-jet cross section analysis from the 2005 RHIC run and
the longitudinal double-spin asymmetry at mid-rapidity for the 2006
data sample are discussed in these proceedings. Projections
on future di-jet measurements at STAR are provided.
\end{abstract}

\date{\today}

\maketitle

\section{Introduction}  


The longitudinal STAR spin physics program profits enormously from the
unique capabilities of the STAR experiment for measuring large
acceptance jet production, identified hadron production and photon
production.  Constraining the polarized gluon distribution function,
$\Delta g$, through inclusive measurements has been, so far, the prime
focus of the physics analysis program of the Run
3/4~\cite{Abelev:2006uq}, Run 5~\cite{Abelev:2007vt} and Run
6~\cite{ref_spin2008_murad} data samples.
%
%
The recent STAR inclusive jet \cite{Abelev:2007vt, ref_spin2008_murad} results along with the PHENIX neutral
pion results \cite{Adare:2008final} have been used for the first time to constrain $\Delta g$
in a NLO global analysis along with semi-inclusive and inclusive DIS
data \cite{deFlorian:2008mr}. The RHIC data sets have been shown to provide strong constraints 
on $\Delta g$ for $0.05<x<0.2$ in this analysis.
%
%
Inclusive measurements, such as inclusive jet production, integrate
over a fairly large $x$ region for a given jet transverse momentum
region. While those measurements provide a strong constraint on the
value of $\Delta g$ integrated over a range in $x$, those measurements do not
permit a direct sensitivity to the actual $x$ dependence. This 
motivates the need for correlation measurements in polarized
proton-proton collisions.

\begin{figure}[t]
\centerline{\includegraphics[width=75mm]{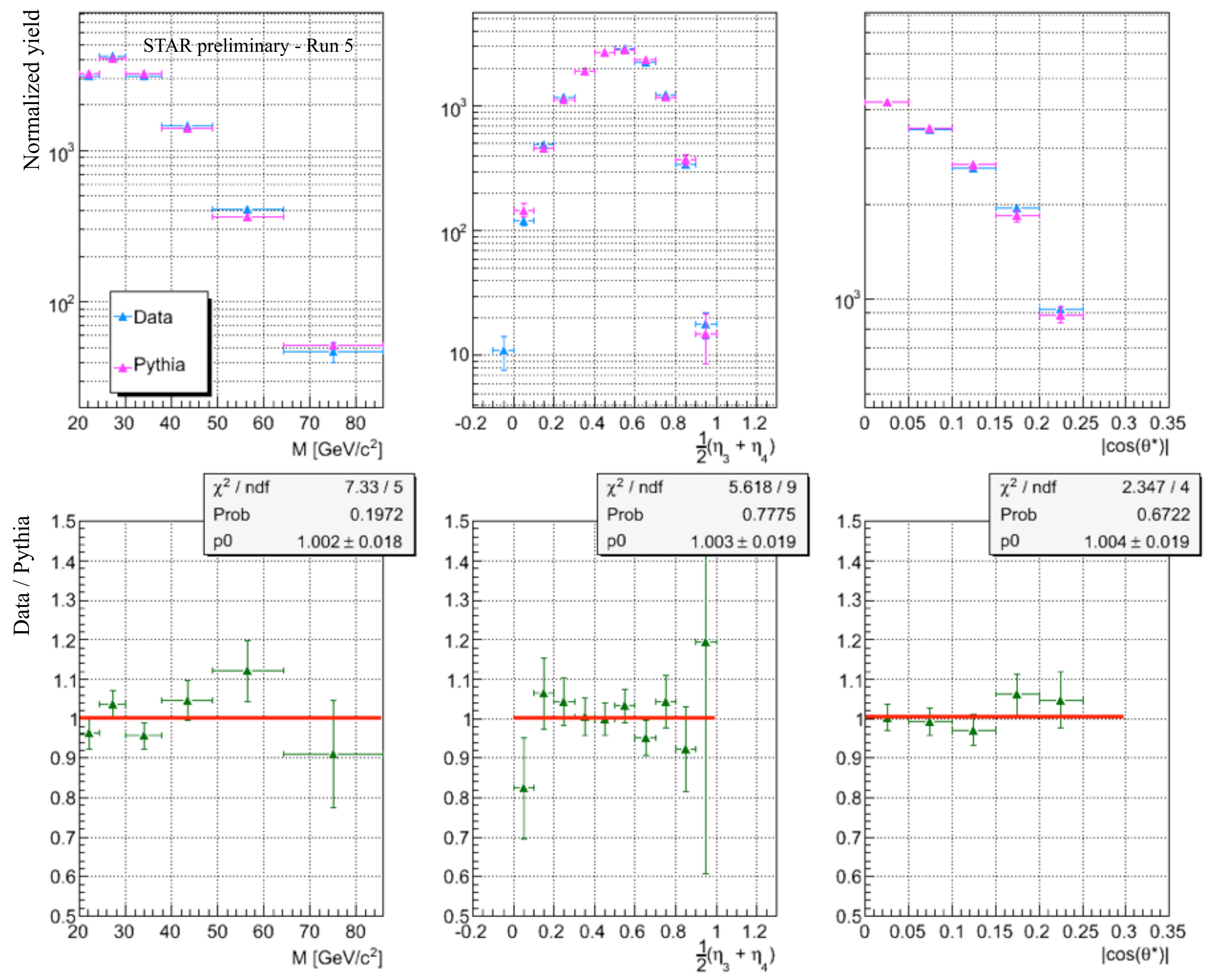}}
\label{fig_1}
\caption{{\it Comparison of data (2005 RHIC run) and a PYTHIA-MC sample for
three di-jet variables, the invariant mass ($M$), the mean of the di-jet pseudo-rapidities ($1/2(\eta_{3}+\eta{4})$) and the cosine of the
center-of-mass scattering angle ($\cos\theta^{*}$).}}
\end{figure}

\section{Status and prospects of di-jet production}


Correlation measurements such as those for di-jet production allow for
a better constraint of the partonic kinematics and thus the shape of
$\Delta g$. At LO, the di-jet invariant mass, $M$, is proportional to
the product of the $x$ values of the partons,
$M=\sqrt{s}\sqrt{x_{1}x_{2}}$, whereas the pseudorapidity sum of the
final-state jets, $\eta_{3}+\eta_{4}$, is proportional to the
logarithm of the ratio of the $x$ values,
$\eta_{3}+\eta_{4}=\ln\left(x_{1}/x_{2}\right)$.  Photon-jet
coincidence measurements are expected to provide a theoretically clean
way to extract $\Delta g$~\cite{Bland:1999}. A LO extraction of $\Delta g$ alone would
allow a model-independent way to constrain the $x$ dependence, which
would be an important contribution, without making an a priori
assumption on the functional form of $\Delta g$ as is currently
required in a global analysis. This has been shown for photon-jet
events in simulations~\cite{Bland:1999}.  The feasibility for a LO extraction of $\Delta
g$ as a function of $x$ for the case of di-jet production still has to
be demonstrated. Measurements at both $\sqrt{s}=200\,$GeV and
$\sqrt{s}=500\,$GeV are preferred to maximize the kinematic reach in
$x$ and possibly provide a means to observe effects of scaling
violations at fixed $x$ by measuring different $p_{T}$ values. The
wide acceptance of the STAR experiment permits reconstruction of
di-jet events with different topological configurations,
i.e. different $\eta_{3}$/$\eta_{4}$ combinations, ranging from
symmetric ($x_{1}=x_{2}$) partonic collisions to asymmetric
($x_{1}<x_{2}$ or $x_{1}>x_{2}$) partonic collisions. This, together
with the variation of the center-of-mass energy, constrains $\Delta g$
over a wide range in $x$ of approximately $\sim 2\cdot 10^{-3} < x <
0.3$ for di-jet and photon-jet events.  The NLO framework for correlation measurements does exist and
therefore those measurements can be used in a global analysis~\cite{deFlorian:1998qp}.



\begin{figure}[t]
\centerline{\includegraphics[width=100mm]{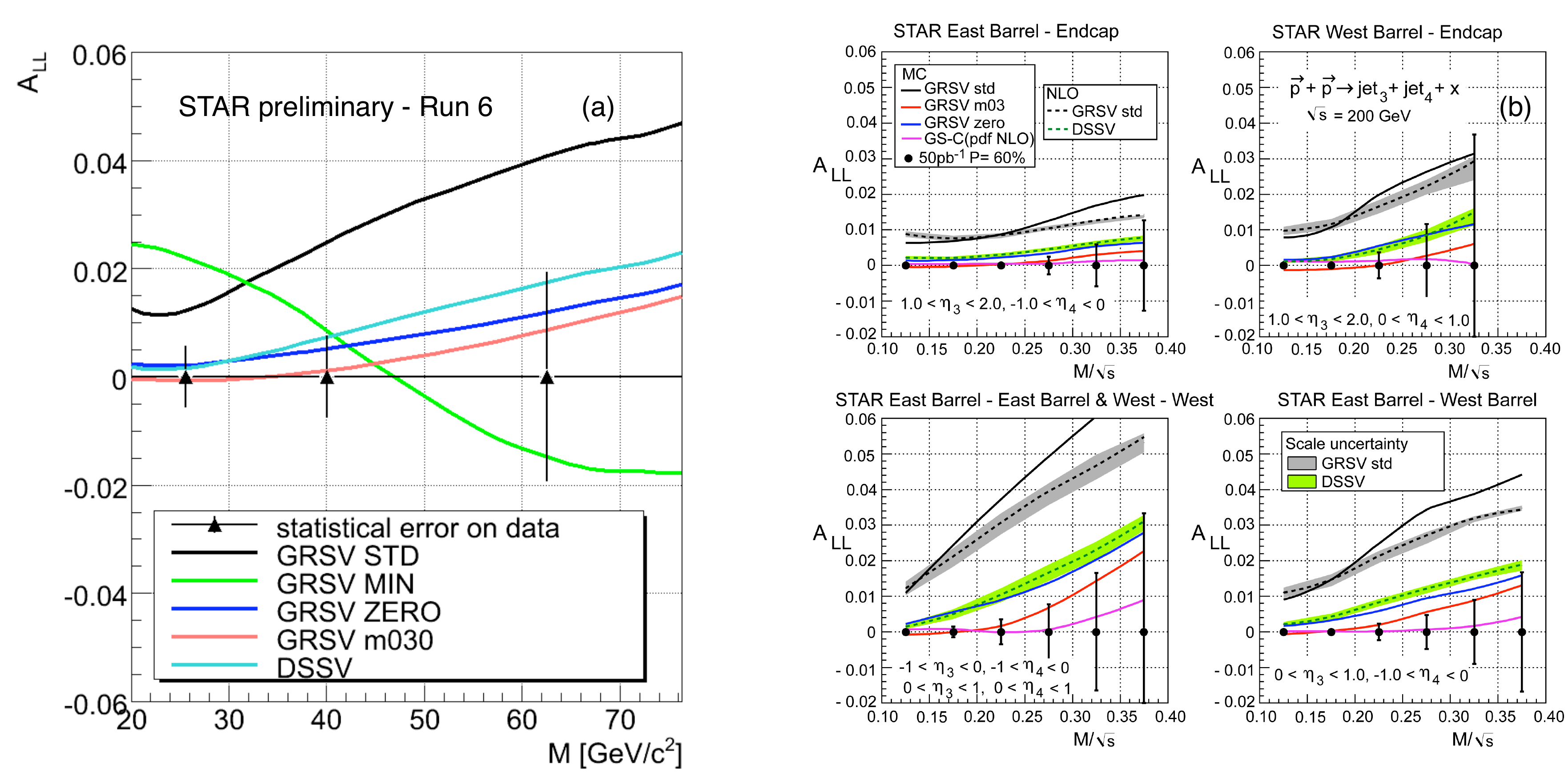}}
\label{fig_2}
\caption{{\it (a) Statistical uncertainty of the longitudinal double-spin 
asymmetry, $A_{LL}$, as a function of the di-jet invariant mass, $M$, for the 2006 RHIC data sample. (b) Statistical precision of the longitudinal 
double-spin asymmetry, $A_{LL}$, for di-jet production as function of the ratio $M/\sqrt{s}$
for different topological combinations of the STAR BEMC and the STAR EEMC acceptance region.}} 
\end{figure}


STAR reconstructs jets with the midpoint cone algorithm using clusters
of charged track momenta measured with the STAR Time Projection Chamber (TPC) and
tower energy deposits in the STAR Barrel Electromagnetic Calorimeter
(BEMC) within a cone radius of $R\equiv\sqrt{\Delta
\eta^{2}+\Delta\phi^{2}}\;$\cite{ref_spin2008_murad}. The jet axis 
was required to be within a fiducial range of
$-0.05<\eta_{\rm{JET}}<0.95\,(-0.7<\eta_{\rm{Detector}}<0.9)$ for the 2005 (2006) data sample
with a cone radius of $R=0.4\,(0.7)$. The dominant fraction of di-jet
events in both the 2005 and 2006 data samples are based on a jet patch
(JP) trigger that required a minimum energy deposition for a group of
towers over a region of $\Delta \eta \times
\Delta \phi = 1.0 \times 1.0$.  This trigger was taken in coincidence
with a minimum-bias condition using the STAR Beam-Beam Counter
(BBC). The di-jet analysis status presented below is based on an
integrated luminosity of approximately $2\,$pb$^{-1}$ and
$5$pb$^{-1}$ for the 2005 and 2006 data samples.


Figure~\ref{fig_1} shows a comparison of data based on the 2005 RHIC
run and a PYTHIA-MC sample for three di-jet variables, the invariant
mass ($M$), the mean of the di-jet pseudorapidities
($1/2(\eta_{3}+\eta{4})$) and the cosine of the center-of-mass
scattering angle ($\cos\theta^{*}$). The top panels show the actual
yield for each variable. The bottom panels display ratios of the data
and MC histograms to quantify the comparison of data and MC
distributions.  The relative normalization has been fixed by the
invariant mass distribution and then applied to both the
$1/2(\eta_{3}+\eta_{4})$ and $\cos\theta^{*}$ distributions, i.e. the
data/MC comparison reflects only one normalization factor. The
comparison is carried out for asymmetric transverse momentum cuts on
both jets using $\rm{min}(p_{T})\geq 7$GeV/c and $\rm{max}(p_{T})\geq
10$GeV/c where $\rm{min}(p_{T})$ ($\rm{max}(p_T)$) refers to the $p_T$
of the jet in the di-jet pair with the smaller (larger) jet
$p_T$. Such an asymmetric requirement on the final-state jet
transverse momenta has been suggested as a means to keep NLO
calculations under control since soft gluon emissions in a
back-to-back jet configuration are not taken into account and would
require resummations. The shape for each di-jet variable in data
and MC is in good agreement. This agreement between data and MC is not
dependent on the asymmetric jet transverse momentum requirement and
even holds for symmetric cuts. A direct comparison of the actual
di-jet cross section to NLO calculations requires further studies such
as the completion of a full evaluation of hadronization and underlying
event effects.


Figure~\ref{fig_2} a) shows the statistical precision of the
longitudinal double-spin asymmetry, $A_{LL}$, as a function of the
di-jet invariant mass, $M$. These uncertainties, extracted from the
2006 data sample, are compared to a LO MC evaluation of $A_{LL}$
computed with a PYTHIA MC sample using different event weights to
account for different polarized gluon distribution functions of
GRSV~\cite{Gluck:2000dy} and DSSV~\cite{deFlorian:2008mr} similar to
the ones discussed in~\cite{ref_spin2008_murad}. The size of the
statistical uncertainty at the highest invariant mass bin is at the
level of the difference between GRSV-STD and DSSV.


Figure~\ref{fig_2} b) shows the statistical precision of the longitudinal double-spin
asymmetry, $A_{LL}$, for di-jet production as a function of
$M/\sqrt{s}$ for different topological combinations of the STAR BEMC
and the STAR Endcap Electromagnetic Calorimeter (EEMC) acceptance
regions. At LO, the ratio $M/\sqrt{s}$ is equal to
$\sqrt{x_{1}x_{2}}$. Taking into account the different $\eta$ ranges
covered, and equivalently, the different $\cos\theta^{*}$ regions
being probed, each panel represents a different range in
$x_{1}$/$x_{2}$. At LO, $\cos\theta^{*}$ amounts to
$\tanh\left(\frac{\eta_{3}-\eta_{4}}{2}\right)$. The upper left panel
effectively probes asymmetric partonic collisions where predominantly
a low-x gluon collides with a high-x quark at large invariant
masses. The effective variation
of $x_{1}$ and $x_{2}$ amounts to $0.2<x_{1}<0.6$ and
$0.07<x_{2}<0.2$. In contrast, a kinematic region of larger $x$ values in $\Delta g$
is probed at predominantly symmetric partonic collisions such as the one
shown in the lower right panel. The effective variation of $x_{1}$ and $x_{2}$ is roughly
equal and given by the horizontal axis of the lower right panel.  The
projected uncertainties are shown for a luminosity of $50\,$pb$^{-1}$
and a beam polarization of $60\%$. Those projected uncertainties are
compared to a LO evaluation of $A_{LL}$ and a full NLO $A_{LL}$
calculation. Scale uncertainties are shown as a shaded band for DSSV
and GRSV-STD reflecting a variation of the invariant mass $M$ as a
hard scale of $2M$ and $0.5M$. Asymmetric cuts are imposed for the LO
MC and the NLO determination of $\rm{min}(p_{T})\geq 7$GeV/c and
$\rm{max}(p_{T})\geq 10$GeV/c.  The result of a LO MC evaluation using
GS-C \cite{Gehrmann:1995ag} for $\Delta g$ is also shown. This
particular choice of $\Delta g$ is reflected in a large positive gluon
polarization at low $x$, a node around $x\sim 0.1$ and a negative
gluon polarization at large $x$ at the initial scale of $4\,$GeV$^{2}$. GS-C is still consistent with the
current inclusive jet results~\cite{ref_spin2008_murad}.  A cone
radius of $R=0.7$ has been used. Good agreement is found between a LO
MC evaluation of $A_{LL}$ and a full NLO calculation.  Scale
uncertainties are found to be small in comparison to the variation of
the chosen polarized gluon distribution functions, in particular, at
large values of $M$. The projected uncertainties are those requested in the STAR Beam Use Proposal for the upcoming
RHIC data taking period at $\sqrt{s}=200\,$GeV in 2009. Di-jet production will play a critical role in the
future, deepening our understanding of $\Delta g$, in particular, by
constraining its shape.

\section{Acknowledgement}

The contribution by Tai Sakuma is acknowledged by the author of this proceedings
contribution.

\vspace*{-0.5cm}

\doingARLO[\bibliographystyle{aipproc}]
          {\ifthenelse{\equal{\AIPcitestyleselect}{num}}
             {\bibliographystyle{arlonum}}
             {\bibliographystyle{arlobib}}
          }
\bibliography{surrow}

\end{document}